\documentclass[floatfix,aps,prb,groupedaddress,showpacs,amsmath,amssymb,reprint]{revtex4-1}
\usepackage{graphicx}
\begin{document}

\title{Lateral manipulation with combined atomic force and scanning tunneling microscopy using CO-terminated tips}

\author{Julian Berwanger}
\email[]{julian.berwanger@ur.de}
\author{Ferdinand Huber}
\author{Fabian Stilp}
\author{Franz J. Giessibl}
\affiliation{Institute of Experimental and Applied Physics, University of Regensburg, 93053 Regensburg, Germany}

\date{\today}

\begin{abstract}
CO-terminated tips currently provide the best spatial resolution obtainable in atomic force microscopy. Due to their chemical inertness, they allow to probe interactions dominated by Pauli repulsion. The small size and inertness of the oxygen front atom yields unprecedented resolution of organic molecules, metal clusters and surfaces. We study the capability of CO-terminated tips to laterally manipulate single iron adatoms on the Cu(111) surface with combined atomic force and scanning tunneling microscopy at $7\,\mathrm{K}$. Furthermore, we find that even a slight asymmetry of the tip results in a distortion of the lateral force field. In addition, the influence of the tilt of the CO tip on the lateral force field is inverted compared to the use of a monoatomic metal tip which we can attribute to the inverted dipole moment of a CO tip with respect to a metal tip. Moreover, we demonstrate atom-by-atom assembly of iron clusters with CO tips while using the high-resolution capability of the CO tips in between to determine the arrangement of the individual iron atoms within the cluster. Additionally, we were able to laterally manipulate single copper and silicon adatoms without changing or losing the CO from the tip's apex.
\end{abstract}

%\pacs{68.37.Ef, 68.37.Ps}

\maketitle
\section{Introduction}
Since Eigler and Schweizer performed the very first atomic positioning experiments using scanning tunneling microscopy (STM) \cite{Schweizer1990}, atomic manipulation for both manipulation and imaging became a widely used technique in the field of scanning probe microscopy \cite{Stroscio1991,Bartels1997,Stroscio2004,Wolter2012,Celotta2014,Kalff2016,Pawlak2017}. 
The ability to precisely position single atoms individually paved the way to e.\,g. create logical operators \cite{Khajetoorians2011} and nano magnets consisting of a single digit count of iron atoms \cite{Khajetoorians2012,Khajetoorians2013a}. Their atomic arrangement is highly relevant for their magnetic properties but remained unclear due to the lack of structural high-resolution capabilities of metal tips.
This became possible by functionalizing the metal tip with a single CO molecule \cite{Bartels1997a}: Using a CO-terminated tip (CO tip) with non-contact atomic force microscopy (nc-AFM) \cite{Albrecht1991} the atomic structure of molecules \cite{Gross2009,Pavlicek2017,Jelinek2017,Extance2018}, 2D materials \cite{Boneschanscher2014,Schulz2018}, ionic surfaces \cite{Ellner2016}, interfacial water \cite{Peng2018a} and also the internal structure of small metal clusters \cite{Emmrich2015e} could be resolved.
However, the combination of lateral manipulation with the high-resolution capability of CO tips is lacking so far. Combining both could in principle be used to e.g. engineer nano magnets by manipulation while knowing their internal structure by imaging them at the same time \cite{Khajetoorians2013a} and to contribute to the debate whether atomic iron chains, hosting majorana fermions, are terminated by a single atom or multiple atoms at their ends \cite{Nadj-Perge2014,Pawlak2016a}. Moreover, this combination is much more time-efficient than using metal tips for the manipulation process and CO tips for imaging, especially, if the manipulation and imaging cycle is repeated several times as done e.g. in Refs. \cite{Mohn2010,Esat2018}. Chemical reactions of molecules could in principle also be driven via manipulation rather than thermal activation and the reactants and products could be imaged before and afterwards with the same CO tip \cite{DeOteyza2013}. 

In this study, we demonstrate lateral manipulation of single iron adatoms using CO tips and study the differences to commonly-used metal tips. We investigate the influence of the tilt of the tip apex on manipulation and are able to explain the reversed manipulation behavior with CO tips compared to metal tips with a simple model based on the tips' dipoles. Finally, we build up iron clusters atom by atom using CO tips and determine the atomic arrangement of the clusters after every manipulation step with the same CO tip. This demonstrates that reproducible lateral manipulation with CO tips is possible without losing the CO tip and while, at the same time, using their high-resolution capability for atomically-resolved imaging.
\section{Experimental Setup}
All experiments were performed with a custom-built combined atomic force and scanning tunneling microscope at $7\,\mathrm{K}$ equipped with a qPlus sensor \cite{Giessibl1998} ($k=1800\,\mathrm{N/m}$, $f_0=26447.5\,\mathrm{Hz}$, $Q=163249$) operating in frequency-modulation mode \cite{Albrecht1991}. To maximize sensitivity to short-range forces, an amplitude of $50\,\mathrm{pm}$ was chosen \cite{Giessibl1999}. An electrochemically etched bulk tungsten tip which was poked repeatedly into the clean Cu(111) sample in order to generate a monoatomic metal tip. To confirm the presence of a monoatomic tip, we performed the carbon monoxide front atom identification (COFI) method \cite{Welker2012}. For this, less than 0.01 monolayers of CO were dosed onto the surface and the tip was scanned in constant height above a single CO molecule. The CO molecule on the surface acts effectively as a probe which images the AFM tip and reveals its geometrical structure in the frequency shift $\Delta f$ image \cite{Emmrich2015e}. After that, the monoatomic metal tip was functionalized with a CO molecule \cite{Bartels1997a}. By using such tips, the hexagonal Cu(111) surface lattice with fcc and hcp hollow sites was resolved in the $\Delta f$ channel \cite{Bartels1997a,Stroscio2004}. Single iron atoms were evaporated onto the cold Cu(111) surface which preferentially adsorb in fcc hollow sites, as those are energetically favorable by $1.9\,\mathrm{meV}$ over hcp hollow sites \cite{Biedermann2006,Negulyaev2009}. The average kinetic energy at $7\,\mathrm{K}$ is given by $3/2k_\mathrm{B}T=0.9\,\mathrm{meV}$, which is small compared to the diffusion barrier from a fcc hollow site to a fcc hollow of $28.5\,\mathrm{meV}$ \cite{Negulyaev2009}. Hence, thermal diffusion can be neglected in our experiments. Each individual iron adatom is surrounded by six equivalent next-neighbor fcc hollow sites (see inset in Fig. \ref{fig1}(a)).
 
The lateral manipulation experiments were performed with various monoatomic metal and CO tips along the six high-symmetry directions \mbox{$\vec{x}\in\left\lbrace\pm\vec{x}_\mathrm{A},\pm\vec{x}_\mathrm{B},\pm\vec{x}_\mathrm{C}\right\rbrace$}. The initial tip-sample distance $z_\mathrm{start}$ was chosen large enough to start with a flat $\Delta f\left(x,z_\mathrm{start}\right)$ profile. Beginning from $z_\mathrm{start}$, consecutive linescans over the iron adatom were performed while decrementing the tip-sample distance by $5\,\mathrm{pm}$ after each linescan. To make sure that the adatom is manipulated in the forward scanning direction, the tip was retracted by $50\,\mathrm{pm}$ when moving backwards. When the atom was manipulated laterally the tip finished the actual linescan and stopped at the lateral starting position, such that the lateral position of the adatom after manipulation can be deduced by observing the tunneling current curve $I\left(x,z\right)$ of the last backward linescan acquired at $z=z_\mathrm{man}+50\,\mathrm{pm}$. To minimize the influence of thermal drift and piezo creep onto the experiment, all lateral manipulation experiments were carried out with drift compensation in $x$-, $y$- and $z$-direction. Doing so, a drift of less than $10\,\mathrm{fm/s}$ is achieved which translates to less than $18\,\mathrm{pm}$ of lateral and vertical drift during the acquisition time of a full data set.
To isolate the interaction between tip and adatom, the $\Delta f$ above the Cu(111) surface is subtracted from each $\Delta f\left(x,z\right)$ curve. Afterwards, following the procedure introduced by Ternes \textit{et al.} \cite{Ternes2008}, the short-range potential $U_\mathrm{SR}$ between tip and adatom was deconvoluted \cite{Sader2004,Sader2017} and, by lateral differentiation, the lateral force $F_{x,\mathrm{SR}}\left(x,z\right)=-\partial U_{\mathrm{SR}}\left(x,z\right)/\partial x$ acting between tip and adatom was determined.
\section{Experimental Results I}
\begin{figure}
\centering
\includegraphics{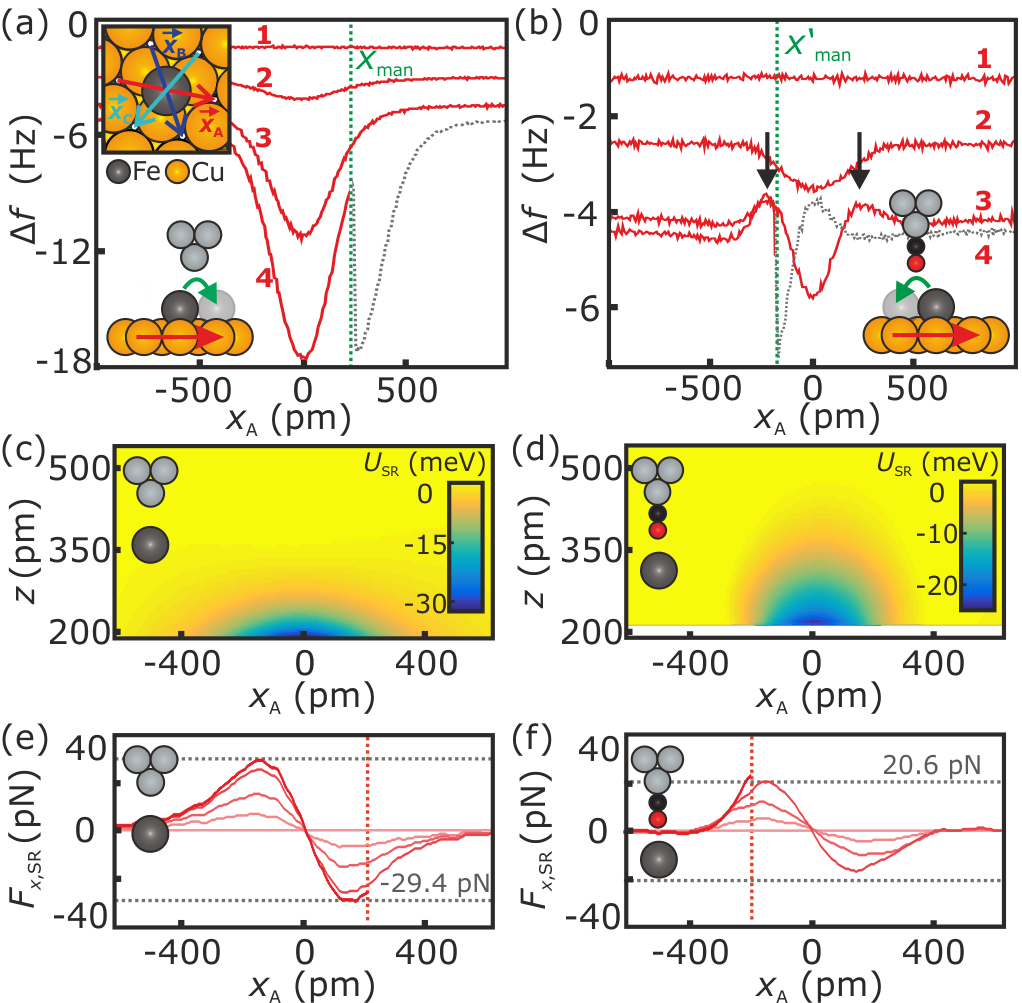}
\caption{(a) Selected raw $\Delta f\left(x_\mathrm{A},z_\mathrm{i}\right)$ traces ($z_\mathrm{1,Me}=571\,\mathrm{pm}$, $z_\mathrm{2,Me}=291\,\mathrm{pm}$, $z_\mathrm{3,Me}=211\,\mathrm{pm}$, $z_\mathrm{4,Me}=186\,\mathrm{pm}$) acquired with a monoatomic metal tip along the $\vec{x}_\mathrm{A}$-direction: The $z$-values refer to the distances to point contact on the Cu(111) surface \cite{Ternes2008,Schneiderbauer2014a} (see also Appendix \ref{AppendixB}). $x_\mathrm{A}=0\,\mathrm{pm}$ indicates the center of the iron adatom before manipulation. It is manipulated at $x_{\mathrm{man}}>0\,\mathrm{pm}$, indicated by the green dotted line. The grey dotted trace is the $\Delta f\left(x_\mathrm{A}\right)$ curve after manipulation. The inset shows a sketch of a Cu(111) surface with an iron adatom adsorbed on a fcc hollow site in top-view. The vectors $\vec{x}_\mathrm{A}$, $\vec{x}_\mathrm{B}$ and $\vec{x}_\mathrm{C}$ indicate the directions to three next-neighbor fcc hollow sites. Panel (b) depicts selected $\Delta f\left(x_\mathrm{A},z_\mathrm{i}\right)$ curves ($z_\mathrm{1,CO}=560\,\mathrm{pm}$, $z_\mathrm{2,CO}=340\,\mathrm{pm}$, $z_\mathrm{3,CO}=220\,\mathrm{pm}$, $z_\mathrm{4,CO}=205\,\mathrm{pm}$) by using a CO tip. The iron adatom is manipulated before the tip has crossed the center of the adatom at $x_{\mathrm{man}}<0\,\mathrm{pm}$. The complete $\Delta f\left(x_\mathrm{A}\right)$ data sets of panels (a) and (b) are shown in Appendix \ref{AppendixA}. (c) Short-range potential $U_\mathrm{SR}\left(x_\mathrm{A},z\right)$ between the metal tip and the iron adatom computed from the $\Delta f\left(x_\mathrm{A}\right)$ data shown in panel (a). The interaction between the adatom and the metal tip is purely attractive. (d) The short-range interaction potential corresponding to the $\Delta f\left(x_\mathrm{A}\right)$ data shown in panel (b) acquired with a CO tip is also entirely attractive. Panels (e) and (f) show selected lines of the corresponding lateral forces $F_{x,\mathrm{SR}}\left(x_\mathrm{A}\right)$. Darker lines refer to closer tip-sample distances. The absolute lateral force increases with decreasing tip-sample distance $z$. Before the deconvolution processes of $\Delta f$, a Gaussian filter ($\sigma=16\,\mathrm{pm}$) was applied.}
\label{fig1}
\end{figure}
Figure \ref{fig1}(a) shows selected $\Delta f\left(x\right)$ linescans acquired with a monoatomic metal tip along the $\vec{x}_\mathrm{A}$-direction. The iron adatom is located at $x_\mathrm{A}=0\,\mathrm{pm}$. At $z_\mathrm{start}=z_\mathrm{1,Me}$, the $\Delta f\left(x_\mathrm{A}\right)$ curve is flat. The $\Delta f\left(x_\mathrm{A},z_\mathrm{2,Me}\right)$ linescan shows an attractive dip centered at the iron adatom. The closer the tip approaches the surface, the deeper the dip above the iron atom evolves. A sharp change in the $\Delta f\left(x_\mathrm{A},z_\mathrm{4,Me}\right)$ curve, indicated by the vertical green dotted line at $x_\mathrm{A}=x_\mathrm{man}$ happens after the tip has passed the center of the iron adatom. After that, the iron atom is imaged at its new lateral position at $x_\mathrm{A}=255\,\mathrm{pm}$ (see Appendix \ref{AppendixA}). The change in $x_\mathrm{A}$ of $255\,\mathrm{pm}$ corresponds to the nearest-neighbor distance of the Cu(111) surface \cite{Bragg1914}. Hence, the iron atom was manipulated laterally to the next-neighbor fcc hollow site along the positive $\vec{x}_\mathrm{A}$-direction (to the right side).

Next, the very same metal tip was functionalized with a CO molecule and the experiment was repeated (see Fig. \ref{fig1}(b)). Curve 2 shows a single attractive dip above the iron adatom. Getting closer results in an evolution of two bumps next to the center, besides of the attractive dip at the center of the iron adatom. In the $x$-$y$-plane, these bumps yield a torus which is well known from previous experiments \cite{Emmrich2015e} (see inset Fig. \ref{fig6}(a)). At the closest tip-sample distance, a sharp change in $\Delta f$ at $x_\mathrm{A}=x'_\mathrm{man}$ occurs before the tip has passed the center of the iron adatom. Here, the iron adatom is manipulated by one lattice position along the negative $\vec{x}_\mathrm{A}$-direction (to the left side).

Figure \ref{fig1}(c) depicts the purely attractive short-range potential $U_\mathrm{SR}\left(x_\mathrm{A},z\right)$ between the metal tip and the iron adatom deconvoluted from the frequency shift $\Delta f$ data shown in Fig. \ref{fig1}(a). The short-range potential $U_\mathrm{SR}\left(x_\mathrm{A},z\right)$ between the CO tip and the iron adatom (see Fig. \ref{fig1}(d)), derived from the $\Delta f$ curves depicted in Fig. \ref{fig1}(b), is also entirely attractive and, furthermore, laterally more confined than its metal tip counterpart due to the high-resolution capability of CO tips. The $z$-values in Figs. \ref{fig1}(c) and (d) refer to the distances to point contact on the Cu(111) surface \cite{Ternes2008,Schneiderbauer2014a} (see Appendix \ref{AppendixB} for further information). 

\begin{figure}
\centering
\includegraphics{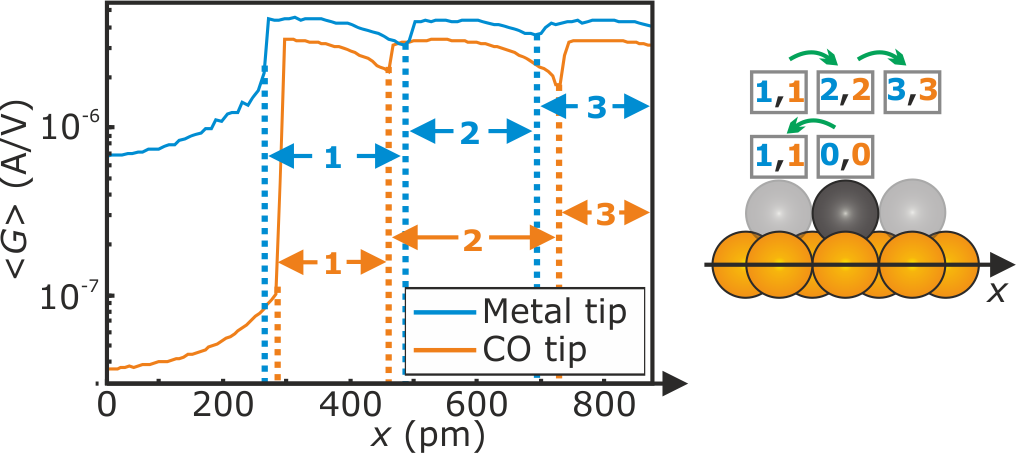}
\caption{Conductance traces $\langle G\rangle$ acquired during two separate lateral manipulation experiments of a single iron adatom using a monoatomic metal tip (blue,top) and a CO tip (orange,bottom) are shown, respectively. The manipulation direction was chosen such that an asymmetric lateral force profile was observed for both tips. The tips' height was reduced until the lateral force threshold for lateral manipulation was overcome on both sides of the adatom. In both cases, the adatom hopped in the first place towards the tip while the tip was approaching the adatom from the left side (lateral position change from \textbf{0} to \textbf{1} at the leftmost vertical blue and orange dotted lines) and afterwards followed the tip moved twice to the right side, indicated by sharp changes in $\langle G\rangle$ marked by vertical dotted lines (lateral position change from \textbf{1} to \textbf{2} and \textbf{2} to \textbf{3}.}
\label{fig2}
\end{figure}

The lateral force curves of the interaction between metal tip and iron adatom and between CO tip and iron adatom, respectively, derived from the deconvoluted short-range potentials $U_\mathrm{SR}\left(x_\mathrm{A},z\right)$ (see Fig. \ref{fig1}(c) and (d)), are qualitatively quite similar (Figs. \ref{fig1}(e) and (f)). The sign of the force value indicates whether the force acting on the adatom points to the left ($+$) or right direction ($-$). Hence, in both cases, the lateral forces acting between tip and adatom are attractive. As the tip approaches the center of the iron adatom from the left side, a positive lateral force acts on the adatom which points to the left side (towards the tip). After the tip has passed the center at $x_\mathrm{A}=0\,\mathrm{pm}$, a negative lateral force acts on the adatom, which points to the right side. As only attractive interactions are observed and the adatom is always manipulated towards the tip, the actual manipulation mode of iron adatoms, using monoatomic metal and CO tips, is pulling, as originally defined by Bartels \textit{et al.} \cite{Bartels1997} (see also Fig. \ref{fig2}). In case of the monoatomic metal tip, the absolute values of the lateral forces are about $1\%$ higher for $x_\mathrm{A}>0\,\mathrm{pm}$ compared to $x_\mathrm{A}<0\,\mathrm{pm}$ and, therefore, the iron adatom is manipulated laterally after the tip has passed the center of the adatom (red dotted vertical line in Fig. \ref{fig1}(e)), since the force threshold \cite{Ternes2008,Emmrich2015d} for lateral manipulation is overcome the first time. In this experiment, the adatom moved only once to the right side. In principle, it would be expected that the adatom would have been manipulated further to the right side, since the force threshold for lateral manipulation is overcome on the right side. A possible reason for not manipulating the adatom further is a slight misalignment of the selected manipulation path or a small variation in the surface potential. Reliable manipulation for longer distances would require further lowering of the tip. The manipulation of the iron adatom with the monoatomic metal tip from Fig. \ref{fig1}(e) but terminated with a CO leads to an inverted asymmetry: The lateral forces before crossing the iron adatom are by about $10\%$ higher than after crossing the adatom (see horizontal grey dotted lines in Fig. \ref{fig1}(f)). Hence, the iron adatom is already manipulated before the tip was passing. After the first lateral manipulation event, the adatom is sitting on the left side of the tip and, hence, the tip is moving away from the adatom. On the right side of the adatom, the lateral force threshold for manipulation is, due to an asymmetry of the lateral force field, not overcome (see Fig. \ref{fig1}(f)) and, therefore, the adatom was not manipulated further.

Again, the adatom can be manipulated in the opposite directions as well, by further approaching the tip to the surface such that the force threshold for lateral manipulation is safely overcome on both sides of the adatom.

In Fig. \ref{fig2} the conductance $\langle G\rangle$ (averaged over the complete oscillation cycle of the sensor) during two lateral manipulation experiments of a single iron adatom along an asymmetric tip direction using a monoatomic metal tip (solid blue curve) and a CO tip (solid orange curve) is shown, respectively. Both tips moved in constant-height across the adatom, starting at $x=0\,\mathrm{pm}$. In both cases, a sharp first change in conductance $\langle G\rangle$ can be detected (marked by the leftmost vertical blue and orange dotted lines), while the conductances showed positive slopes, and, hence the tips were still approaching the center of the adatoms. Since the manipulation events occurred while approaching the center of the atom and both tips laterally manipulate iron adatoms in an attractive mode, the adatoms are in the first place manipulated laterally by one lattice site towards the tips (to the left side). After that, the adatoms are imaged at their new lateral positions (see blue section \textbf{1} for monoatomic metal tip and orange section \textbf{1} for CO tip, respectively). Next, another sharp change in the conductances occurs, while $\partial\langle G\left(x\right)\rangle/\partial x<0$, and afterwards the adatoms are imaged in their new lateral positions (see blue section \textbf{2} for monoatomic metal tip and orange section \textbf{2} for CO tip, respectively). Hence, the adatoms are laterally manipulated, after the tip has passed the center of the iron adatoms, to the right side, to their initial adsorption position. Eventually, another lateral manipulation event occurred for both tips, since another sharp change in the conductances is observed, while $\partial\langle G\left(x\right)\rangle/\partial x<0$. The adatoms are manipulated to the right side by one lattice position and are imaged further at this position (see blue section \textbf{3} for monoatomic metal tip and orange section \textbf{3} for CO tip, respectively). The lateral movements of the tips were stopped at $x=800\,\mathrm{pm}$ in both cases. By comparing both conductance curves with Fig. 2(a) of Ref. \cite{Bartels1997}, an unambiguously identification of the manipulation mode as pulling, can be deduced.

\begin{figure}
\centering
\includegraphics{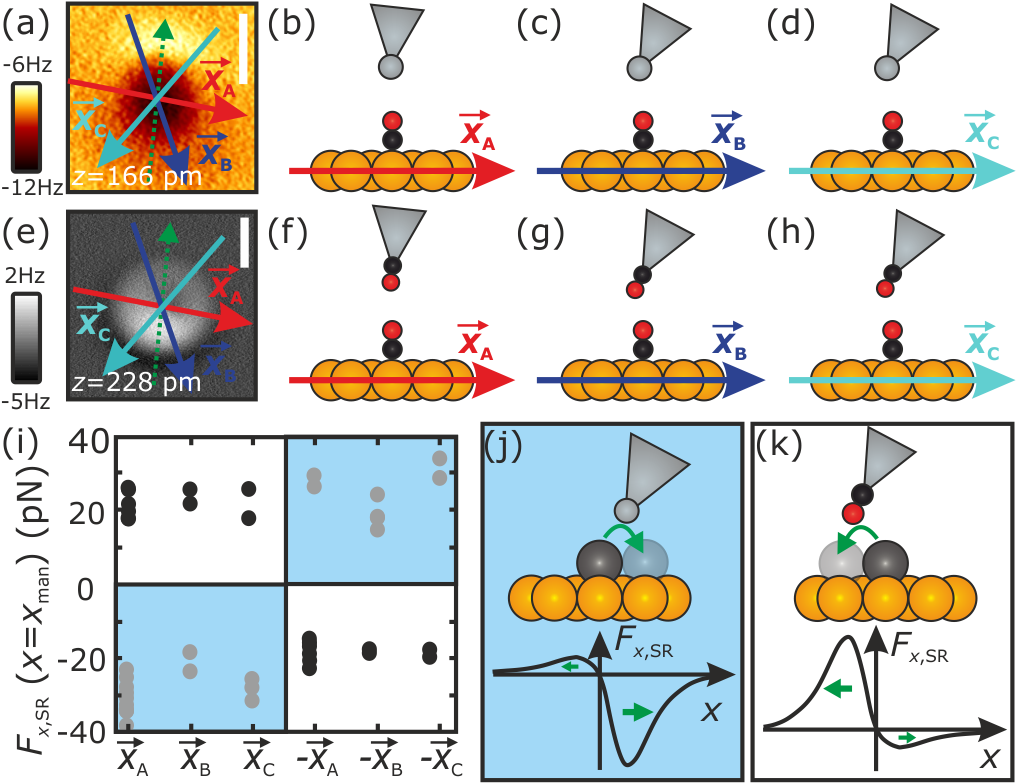}
\caption{(a) Constant-height $\Delta f$ image acquired above a CO molecule (COFI image) shows a tilted monoatomic metal tip \cite{Welker2012,Emmrich2015d}. The scale bars in panels (a) and (b) correspond to $300\,\mathrm{pm}$. Panels (b)-(d) depict the tilt of the monoatomic metal tip along the three high-symmetry directions of the Cu(111) surface. Panel (e) displays the COFI image of the monoatomic metal tip from (a) with a CO molecule attached to its apex. Due to symmetry reasons, the CO molecule adsorbs to the tip's apex as illustrated in panels (f)-(h). (i) Lateral force acting at the moment of manipulation against manipulation direction. The tilt of the tips introduces an asymmetry in the force profiles. In case of the monoatomic metal tip, the iron adatom is at first manipulated when the tilted part of the tip is facing away from the adatom (as sketched in panel (j)). For CO tips, the initial lateral manipulation of the iron adatom takes place when the tilted part of the tip is facing towards the adatom (as sketched in panel (k)). In both cases, the adatom can be manipulated in the opposite directions as well by reducing the tip-sample distance further (see Fig. \ref{fig2}).}
\label{fig3}
\end{figure}

To further investigate the directional dependences of the lateral manipulation, the manipulation experiment was performed in all six high-symmetry directions of Cu(111) and the COFI images of the monoatomic metal and CO tip are analyzed in detail. Figure \ref{fig3}(a) shows the COFI image ($\Delta f\left(x,y\right)$) of the monoatomic metal tip. The image reveals a tilt of the tip whose direction can be determined by investigating the bright sickle in the top part of the image \cite{Welker2012,Emmrich2015d}. This sickle is present because the tilted part of the tip at these $\left(x,y\right)$-positions is closer to the CO molecule and the interaction is already in the repulsive regime compared to the positions mirrored at the center of the CO molecule. To clarify this, the tilt of the metal tip is sketched in Figs. \ref{fig3}(b)-(d) along the high-symmetry manipulation directions. The $\vec{x}_\mathrm{A}$-direction is almost perpendicular to the symmetry axis of the tip (see dashed arrow in Fig. \ref{fig3}(a)) and, therefore, the $F_{x,\mathrm{SR}}\left(x_\mathrm{A}\right)$ curves in Fig. \ref{fig1}(e) show only a slight asymmetry with respect to $x_\mathrm{A}=0\,\mathrm{pm}$. Along the $\vec{x}_\mathrm{B}$- and $\vec{x}_\mathrm{C}$-directions, the tilt of the tip is more significant and, hence, also the lateral force profiles are more asymmetric (see Appendix \ref{AppendixC}). Figure \ref{fig3}(e) shows the COFI image of the monoatomic metal tip from Fig. \ref{fig3}(a) with a CO molecule attached to its apex. Due to symmetry reasons, the CO molecule adsorbs to the tip's apex as sketched in Figs. \ref{fig3}(f)-(h). This suggestion is supported by the observation of the attractive sickle in the bottom part of the image (for a deeper analysis regarding the tilt determination see Appendix \ref{AppendixD} ).

Figure \ref{fig3}(i) depicts the lateral forces acting in the moment of lateral manipulation with respect to the manipulation direction $\vec{x}$. Grey colored data points correspond to manipulation experiments with two monoatomic metal tips whereas black data points result from experiments with three different CO-terminated tips. As discussed before, positive or negative force values indicate whether the iron adatom is manipulated before or after the tip has passed the center of the iron adatom, respectively. Although the metal tip and the CO tip are tilted in the same direction, the influence of the tilt onto the initial manipulation direction is reversed: For monoatomic metal tips, the lateral force is always higher in absolute value when the tilted part of the tip is facing away from the adatom, as sketched in Fig. \ref{fig3}(j), and, therefore, the manipulation occurs at first when the tilted part of the tip is facing away from the adatom. Hence, the forces in Fig. \ref{fig3}(i) are negative for $\vec{x}\in\left\lbrace \vec{x}_\mathrm{A},\vec{x}_\mathrm{B},\vec{x}_\mathrm{C}\right\rbrace$ and positive for $\vec{x}\in\left\lbrace -\vec{x}_\mathrm{A},-\vec{x}_\mathrm{B},-\vec{x}_\mathrm{C}\right\rbrace$. Contrary behavior is observed for the manipulation with CO tips: The absolute value of the lateral force shows a maximum when the tilted part of the tip is facing the adatom, as sketched in Fig. \ref{fig3}(k) and, thus, the lateral forces are positive for $\vec{x}\in\left\lbrace \vec{x}_\mathrm{A},\vec{x}_\mathrm{B},\vec{x}_\mathrm{C}\right\rbrace$ and negative for $\vec{x}\in\left\lbrace -\vec{x}_\mathrm{A},-\vec{x}_\mathrm{B},-\vec{x}_\mathrm{C}\right\rbrace$, respectively. Therefore, attaching a CO molecule to a monoatomic metal tip inverts the influence on the asymmetry in the lateral force profile. As a consequence, a metal tip oriented as in Fig. \ref{fig3}(j) needs to be closer to the surface for a manipulation to the left than for a manipulation to the right and vice-versa for a CO tip like in Fig. \ref{fig3}(k).

To investigate the reason for this inversion, we applied an analytic model considering van der Waals (vdW) \cite{Lennard-Jones1931} and electrostatic (ES) \citep{DeCoulomb1788} interactions between a tilted, two-dimensional tip and an iron atom, which will be explained in the following.

\section{Analytical Model}

\begin{figure}
\centering
\includegraphics{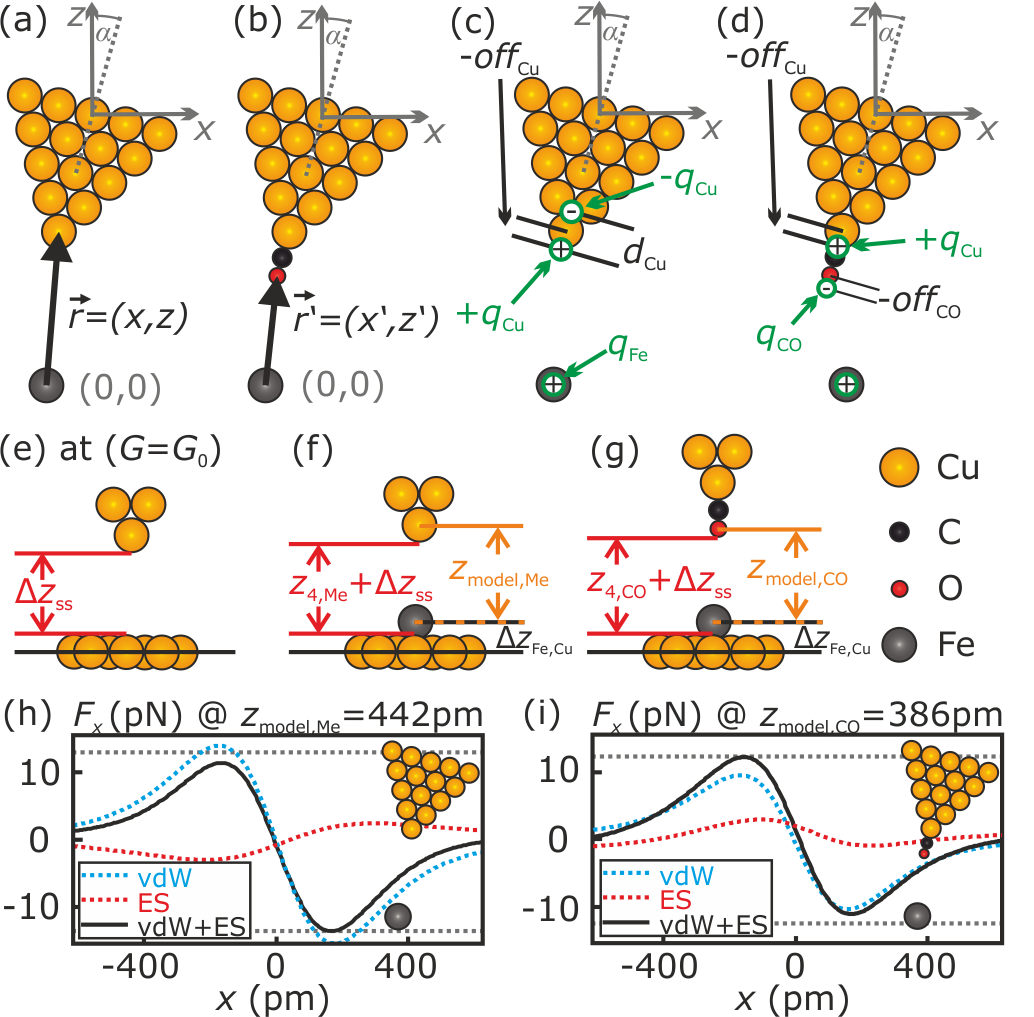}
\caption{Sketch of the (a) monoatomic metal tip and the (b) CO tip used for the analytical model. Copper atoms are used to model the metal tip since the tip was repeatedly poked into the clean Cu(111) surface during tip shaping. Panels (c) and (d) depict the positions of the point charges of the monoatomic metal tip and the CO tip. The parameters were set to: $q_\mathrm{Cu}=\pm +0.13 e$ \cite{Schneiderbauer2014a}, $d_\mathrm{Cu}=135\,\mathrm{pm}$ \cite{Schneiderbauer2014a}, $\textit{off}_\mathrm{Cu}=0\,\mathrm{pm}$ \cite{Schneiderbauer2014a},  $q_\mathrm{CO}=\pm -0.03 e$ \cite{Schneiderbauer2014a}, $\textit{off}_\mathrm{CO}=-100\,\mathrm{pm}$ \cite{Ellner2016} and $q_\mathrm{Fe}=\pm +0.1 e$. (e) A monoatomic tip experience a conductance of $G=G_0=\frac{2e^2}{h}$ at a shell-shell-distance between the tip's apex atom and an atom of the topmost layer of the Cu(111) surface of $\Delta z_\mathrm{ss}=172\,\mathrm{pm}$ \cite{Gustafsson2017}. (f) The core-core distance between iron adatom and the topmost copper layer ($\Delta z_\mathrm{Fe,Cu}=186\,\mathrm{pm}$) \cite{Polesya2018}, the atomic radius of copper ($135\,\mathrm{pm}$) and the closest approach in experiment $z_\mathrm{4,Me}$ (see Fig. \ref{fig1}(a)) yield the closest core-core distance between the apex atom and the iron adatom in experiment of: $z_\mathrm{model,Me}=442\,\mathrm{pm}$. (g) The same considerations for the CO tip, with the atomic radius of oxygen ($60\,\mathrm{pm}$), the experimental closest approach $z_\mathrm{4,CO}$ (see Fig. \ref{fig1}(b)) and obeying the $z$-distance relation between monoatomic metal tips and CO-functionalized monoatomic metal tips \cite{Schneiderbauer2014a} (Appendix \ref{AppendixB}), result in a core-core distance between the tip's oxygen atom and the iron adatom of $z_\mathrm{model,CO}=386\,\mathrm{pm}$. The lateral force $F_x$ acting due to pure vdW (blue dotted curve), pure ES (red dotted curve) and vdW+ES interaction (black solid curve) between the iron atom and (h) the monoatomic metal tip and (i) the CO tip, respectively ($\alpha=20^\circ$).}
\label{fig4}
\end{figure}

\subsection{Van der Waals interaction}
The vdW interaction is purely attractive. For a tip consisting of $N$ atoms, the vdW potential between a single iron atom and the tip can be written as:
\begin{equation}
U_\mathrm{vdW}=-\sum_{i=1}^{N}\frac{C6_{spec(i),\mathrm{Fe}}}{r^6_{i,\mathrm{Fe}}}
\end{equation}
Here, the parameter $C6_{spec(i),\mathrm{Fe}}$ can be calculated for various atomic species $spec(i)$ \cite{Gould2016} (see Table \ref{Table1}).
 \begin{table}
 \caption{$C6_k$ coefficients in dependence of interaction partner $k$ \cite{Gould2016}\label{Table1}}
 \begin{ruledtabular}
 \begin{tabular}{l|ccccc}
 Interaction partner $k$&Cu,Fe&C,Fe&O,Fe&Cu,C&Cu,O \\
 $C6_{k} [10^{-77}\,\mathrm{Jm^6}]$&$3.64$&$1.42$&$0.73$&$0.99$&$0.50$\\
 \end{tabular}
 \end{ruledtabular}
 \end{table}
The value $r_{i,\mathrm{Fe}}=\left|\vec{r}_{i,\mathrm{Fe}}\right|=\sqrt{\left( x_{i,\mathrm{Fe}}\right) ^2+\left( z_{i,\mathrm{Fe}}\right) ^2}$ describes the distance between the centers of the interacting atoms. Figure \ref{fig4}(a) and (b) depict the tip models used for the calculations: The monoatomic metal tip consists of $15$ copper atoms which are ordered in a pyramidal shape. The distance between the copper atoms is set to the lattice constant of copper ($a_\mathrm{Cu,NN}=361\,\mathrm{pm}$). The CO tip consists of the same metal background while a CO molecule is added additionally to the metal tip’s apex atom. The distance of the center of the oxygen atom to the center of the copper atom of the apex of the metal background is given by: $r_\mathrm{O}+2∙r_\mathrm{C}+r_\mathrm{Cu}=60\,\mathrm{pm}+2\cdot 70\,\mathrm{pm}+135\,\mathrm{pm}=335\,\mathrm{pm}$, while $r_i$ with $i=\lbrace\mathrm{O},\mathrm{C},\mathrm{Cu}\rbrace$ are the atomic radii of the different atomic species. The tips can be tilted by an angle $\alpha$.

\subsection{Electrostatic interaction}
Due to the Smoluchowski effect, a metal tip possesses a positive dipole at its apex \cite{Smoluchowski1941}. Since all experiments were performed at very close tip-sample distances, the dipole of the tip was modeled by two single charges $q_\mathrm{Cu}=\pm 0.13\, e$ with opposite sign, separated by $d_\mathrm{Cu}=135\,\mathrm{pm}$ \cite{Schneiderbauer2014a} (see Fig. \ref{fig4}(c)). The CO tip’s dipole is inverted \cite{Schneiderbauer2014a,Ellner2016}: In case of the CO tip, a negative charge $q_\mathrm{CO}=-0.03\,e$ \cite{Schneiderbauer2014a} was put with a slight offset of $\textit{off}_\mathrm{CO}=-100\,\mathrm{pm}$ \cite{Ellner2016} on the oxygen atom of the tip and the positive charge of $q_\mathrm{Cu}=+0.13\, e$ is kept in the center of the front most atom of the metal background (see Fig. \ref{fig4}(d)). In both cases, a positive charge of $q_\mathrm{Fe}=+0.10\, e$ was placed in the center of the iron adatom on the surface. The electrostatic interaction between tip and iron adatom can be written as the Coulomb interaction:
\begin{equation}
U_\mathrm{ES}=+\frac{1}{4\pi\epsilon_0}\sum_{i=1}^{N}\frac{q_\mathrm{Fe}q_\mathrm{spec(i)}}{r_{i,\mathrm{Fe}}}
\end{equation}

\subsection{Aligning $z$ between model and experiment}
To compare the analytical model with the experimental data, the $z$-axes need to be matched. In experiment, the simultaneously acquired conductance $G$ can be used to determine absolute distances. Gustafsson \textit{et al.} found that a monoatomic tip experiences a conductance of $G=G_0=\frac{2e^2}{h}$ when the shell of the tip's apex is still separated by $\Delta z_\mathrm{ss}=172\,\mathrm{pm}$ from the shell of the topmost layer of the Cu(111) surface \cite{Gustafsson2017} (see Fig. \ref{fig4}(e)). Taking the core-core distance between the iron adatom and the topmost copper layer of $\Delta z_\mathrm{Fe,Cu}=186\,\mathrm{pm}$, derived by density functional theory \cite{Polesya2018}, the atomic radius of copper ($135\,\mathrm{pm}$) and the closest approach in experiment $z_\mathrm{4,Me}$ (see Fig. \ref{fig1}(a)) into account, the core-core distance between the frontmost atom of the monoatomic metal tip and the iron adatom can be derived: $z_\mathrm{model,Me}=442\,\mathrm{pm}$ (see Fig. \ref{fig4}(f)). Performing the same considerations for the CO tip, but using the atomic radius of oxygen ($60\,\mathrm{pm}$), the closest approach in experiment $z_\mathrm{4,CO}$ (see Fig. \ref{fig1}(b)) and obeying the $z$-distance relation between monoatomic metal tips and CO-functionalized monoatomic metal tips \cite{Schneiderbauer2014a} (Appendix \ref{AppendixB}), the core-core distance between the oxygen atom of the tip and the iron adatom for the experimental closest approach can be determined: $z_\mathrm{model,CO}=386\,\mathrm{pm}$ (see Fig. \ref{fig4}(g)).

Figure \ref{fig4}(h) and (i) shows the resulting lateral forces acting between the model tips and the iron atom. The iron atom is located at $x=0\,\mathrm{pm}$. Pure vdW interaction between both model tips and the iron atom (blue dotted curves) yield the same asymmetry of the lateral force curves while the lateral forces are higher in absolute value when the tilted part of the tips is facing away from the iron atom ($\left|\mathrm{max}\left(F_x\right)\right|<\left|\mathrm{min}\left(F_x\right)\right|$). By including the ES interaction (red dotted curve), the asymmetry for the monoatomic metal tip stays the same (solid black curve in Fig. \ref{fig4}(h)). However, the reversed dipole of the CO tip yields an inversion of the asymmetry (see Fig. \ref{fig4}(i)). Hence, this analytical model consisting of vdW and ES interactions describes qualitatively the experimental observations (see Fig. \ref{fig3}(j) and (k)). Interestingly, the maxima of the absolute value of the lateral forces of the model $\mathrm{max}\left(\left|F_x\right|\right)$ are quite similar for the monoatomic metal tip ($\left|-13.6\,\mathrm{pN}\right|$) and the CO tip ($11.4\,\mathrm{pN}$). Compared to the experimental values of $27\pm 6\,\mathrm{pN}$ (monoatomic metal tip) and $20\pm 4\,\mathrm{pN}$ (CO tip) the lateral forces of the analytical model are by about $60\%$ (monoatomic metal tip) and $50\%$ (CO tip) too low. This can be explained by the two-dimensional tip in the model instead of a three-dimensional tip as in experiment.

\begin{figure*}
\centering
\includegraphics{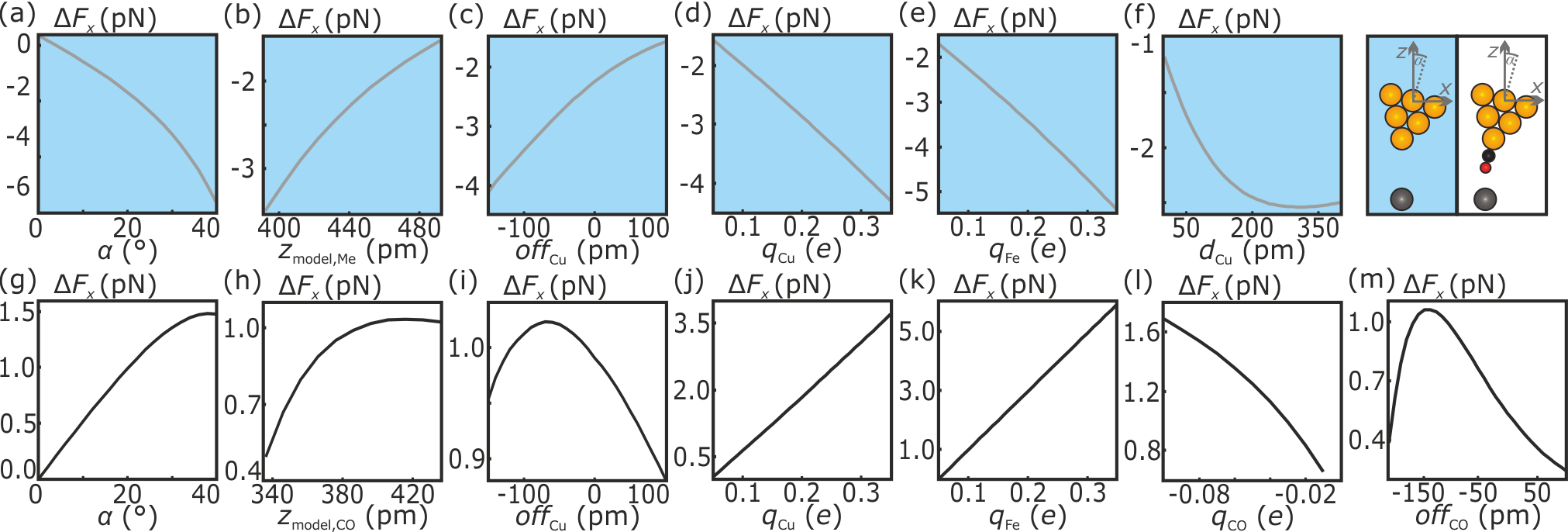}
\caption{The sum of the two extrema of the lateral force curve $\Delta F_x=\mathrm{max}\left(F_x\right)+\mathrm{min}\left(F_x\right)$ is plotted for the vdW + ES model as a function of the parameters (a) $\alpha$, (b) $z_\mathrm{model,Me}$, (c) $\textit{off}_\mathrm{Cu}$, (d) $q_\mathrm{Cu}$, (e) $q_\mathrm{Fe}$ and (f) $d_\mathrm{Cu}$ in case of the monoatomic metal tip. The influence of $\alpha$, $z_\mathrm{model,CO}$, $\textit{off}_\mathrm{Cu}$, $q_\mathrm{Cu}$, $q_\mathrm{Fe}$, $q_\mathrm{CO}$ and $\textit{off}_\mathrm{CO}$ onto the sum of the two extrema of the lateral force curve $\Delta F_x=\mathrm{max}\left(F_x\right)+\mathrm{min}\left(F_x\right)$ is depicted in panels (g)-(m) in case of the CO tip. The constant parameters were set to $d_\mathrm{Cu}=135\,\mathrm{pm}$ \cite{Schneiderbauer2014a}, $\textit{off}_\mathrm{Cu}=0\,\mathrm{pm}$ \cite{Schneiderbauer2014a}, $\textit{off}_\mathrm{CO}=-100\,\mathrm{pm}$ \cite{Ellner2016}, $z_\mathrm{model,Me}=442\,\mathrm{pm}$, $z_\mathrm{model,CO}=386\,\mathrm{pm}$ and $\alpha=20^\circ$, respectively.}
\label{fig5}
\end{figure*}

To determine the robustness of the model, the dependencies of the model on the starting parameters are investigated in the following. For this purpose, the sum of the two extrema of the lateral force curve $\Delta F_x=\mathrm{max}\left(F_x\right)+\mathrm{min}\left(F_x\right)$ is plotted against the parameters $\alpha$, $z_\mathrm{model,Cu}$, $z_\mathrm{model,CO}$, $\textit{off}_\mathrm{Cu}$, $q_\mathrm{Cu}$, $q_\mathrm{Fe}$, $d_\mathrm{Cu}$, $q_\mathrm{CO}$ and $\textit{off}_\mathrm{CO}$ (see Fig. \ref{fig5}). Hence, a positive sign indicates that the lateral force is higher in absolute value when the tilted part of the tip is facing towards the iron atom ($\left|\mathrm{max}\left(F_x\right)\right|>\left|\mathrm{min}\left(F_x\right)\right|$) while a negative sign indicates that the lateral force is higher in absolute value when the tilted part of the tip is facing away from the iron atom ($\left|\mathrm{max}\left(F_x\right)\right|<\left|\mathrm{min}\left(F_x\right)\right|$).
In case of the monoatomic metal tip, $\Delta F_x$ is always negative for various tilt angles of the tip $\alpha=\left[0^\circ,40^\circ\right]$, various $z$-positions $z_\mathrm{model,Cu}=[442\,\mathrm{pm}\pm 50\,\mathrm{pm}]$, various offsets $\textit{off}_\mathrm{Cu}=[-150\,\mathrm{pm},150\,\mathrm{pm}]$ of the positive charge of the front most tip atom, different charges on the tip's apex atom and the iron adatom $q_\mathrm{Cu},q_\mathrm{Fe}=[0.05\,e,0.35\,e]$ and various distances $d_\mathrm{Cu}=[0\,\mathrm{pm},400\,\mathrm{pm}]$ between the positive and negative charges $q_\mathrm{Cu}=\pm 0.13\, e$ (see Figs. \ref{fig5}(a)-(f)). In case of the CO tip, $\Delta F_x$ is always positive for the same tilt angle range of the tip $\alpha=\left[0^\circ,40^\circ\right]$, various $z$-positions $z_\mathrm{model,Cu}=[386\,\mathrm{pm}\pm 50\,\mathrm{pm}]$, the same offsets $\textit{off}_\mathrm{Cu}=[-150\,\mathrm{pm},150\,\mathrm{pm}]$ of the positive charge of the front most tip atom $q_\mathrm{Cu}=\pm 0.13\, e$, different charges on the metal tip's apex atom and the iron adatom $q_\mathrm{Cu},q_\mathrm{Fe}=[0.05\,e,0.35\,e]$, various charges on the oxygen atom $q_\mathrm{CO}=[-0.10\,e,-0.01\,e]$ and for various offsets $\textit{off}_\mathrm{CO}=[-150\,\mathrm{pm},150\,\mathrm{pm}]$ of the negative charge of the oxygen atom of the tip $q_\mathrm{CO}=-0.03\, e$ (see Figs. \ref{fig5}(g)-(m)). The fact that the model qualitatively reproduces the inverted asymmetry of the lateral force field using monoatomic metal tips and CO tips even for different simulation parameters is a proof of the robustness of the analytical model.

\section{Experimental Results II}
\begin{figure*}
\centering
\includegraphics{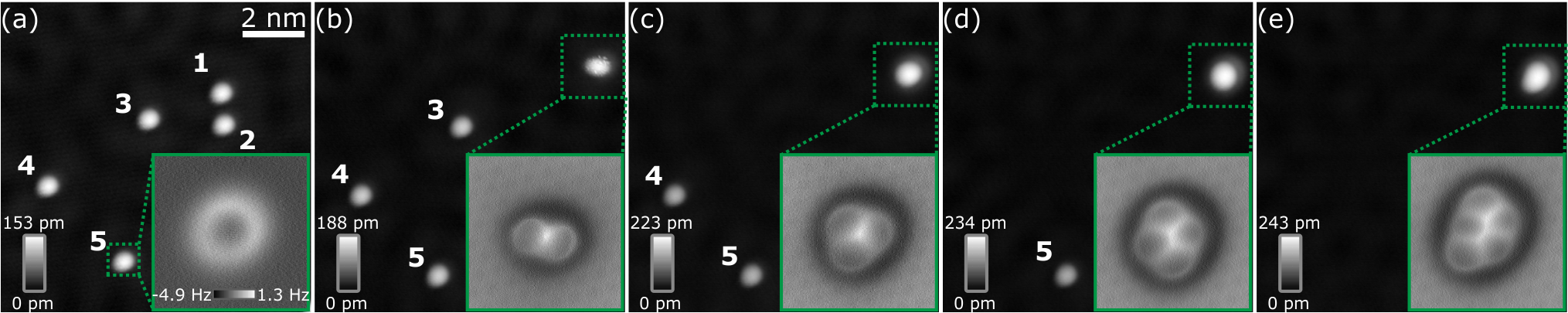}
\caption{Panels (a) through (e) show topographic images of a sample area with five iron atoms acquired in constant-current mode ($V_\mathrm{tip}=-10\,\mathrm{mV}$, $\langle I\rangle=10\,\mathrm{pA}$). By performing controlled lateral manipulation using a CO tip, a cluster consisting of five individual iron adatoms is formed atom by atom. The cluster is atomically resolved after each lateral manipulation event: The insets of each panel depict the $\Delta f$ image acquired in (a) constant-height mode and (b)-(e) constant-current mode, respectively ($V_\mathrm{tip}=-10\,\mathrm{mV}$, $\langle I\rangle=300\,\mathrm{pA}$, same color scales).}
\label{fig6}
\end{figure*}

The average lateral force threshold needed for lateral manipulation of an iron adatom on Cu(111) using monoatomic metal tips is $27\pm 6\,\mathrm{pN}$ which is comparable to the value of $20\pm 4\,\mathrm{pN}$ that holds for CO tips (see Fig. \ref{fig3}(i)). Moreover, our determined force values are similar to the lateral force needed to manipulate a cobalt adatom laterally with an uncharacterized metal tip on Cu(111) ($17\pm3\,\mathrm{pN}$) \cite{Ternes2008}. However, Negulyaev \textit{et al.} experimentally found a diffusion barrier of $22\pm7\,\mathrm{meV}$ for iron adatoms on Cu(111) which matches their calculated diffusion barrier of $28.5\,\mathrm{meV}$ \cite{Negulyaev2009}. By using a sinusoidal model potential, this translates into a lateral force needed to overcome these barriers of $65\pm20\,\mathrm{pN}$ and $84.4\,\mathrm{pN}$, respectively. Emmrich \textit{et al.} \cite{Emmrich2015d} proposed a lowering of the potential barrier by $50\%$ due to the presence of the AFM tip in lateral manipulation experiments of single CO molecules on Cu(111) and recent experiments report a bond weakening between CO and Cu(111) induced by the presence of the tip \cite{Okabayashi2018}. In our experiments with single iron adatoms on Cu(111), the lowering of the potential barrier is up to $70\%$ with respect to the theoretical value of $28.5\,\mathrm{meV}$ \cite{Negulyaev2009}, and, hence, even higher.

\begin{figure*}
\centering
\includegraphics{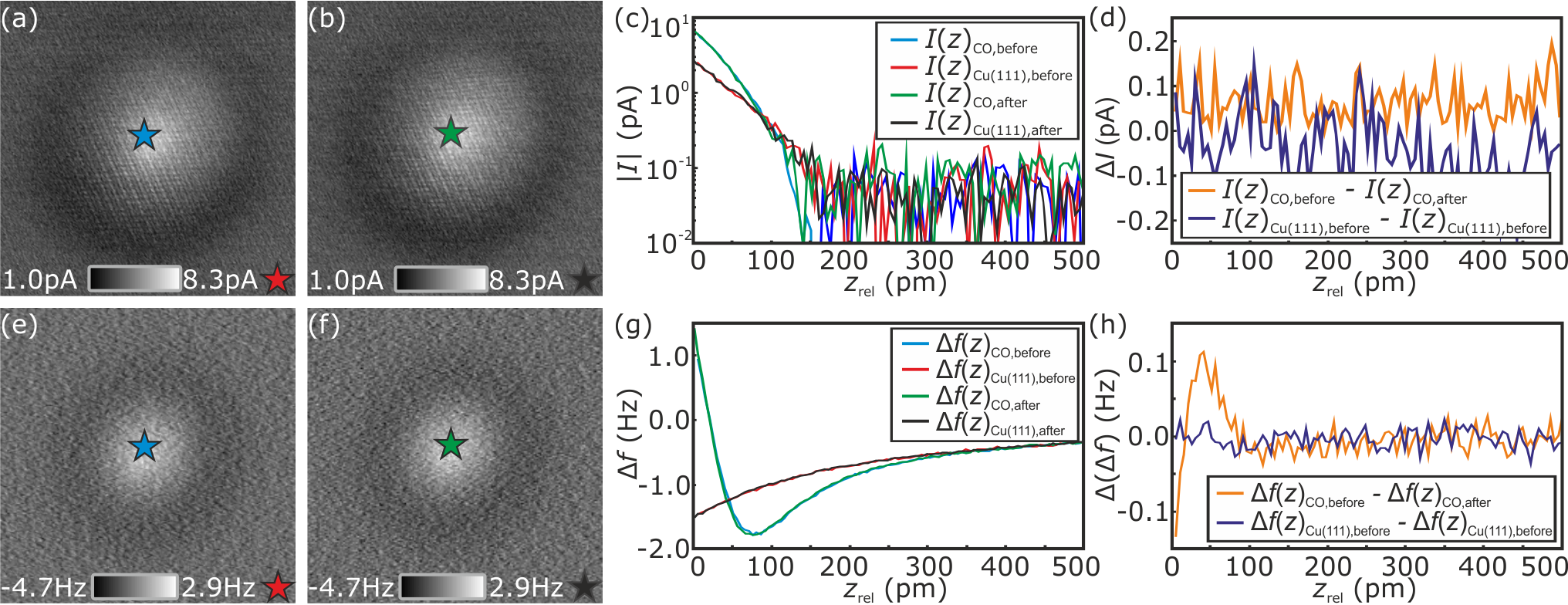}
\caption{Panels (a)\&(e) and (b)\&(f) show simultaneously acquired $I\left(z=\mathrm{const.}\right)$ and $\Delta f\left(z=\mathrm{const.}\right)$ images before and after the assembly of the pentamer shown in Fig. \ref{fig6}. In panel (c) the $I\left(z\right)$ curves on the spots marked in panels (a)\&(b) are depicted. (d) The differences $I\left(z\right)_\mathrm{CO,before}-I\left(z\right)_\mathrm{CO,after}$ and $I\left(z\right)_\mathrm{Cu(111),before}-I\left(z\right)_\mathrm{Cu(111),after}$ are within the noise level of the STM amplifier. (h) The differences $\Delta f\left(z\right)_\mathrm{CO,before}-\Delta f\left(z\right)_\mathrm{CO,after}$ and $\Delta f\left(z\right)_\mathrm{Cu(111),before}-\Delta f\left(z\right)_\mathrm{Cu(111),after}$ of the $\Delta f\left(z\right)$ curves depicted in panel (g) show only a small deviation of less than $0.12\,\mathrm{Hz}$ which can be attributed to a tiny offset in the tips' lateral positions of the before and after spectra with respect to each other.}
\label{fig7}
\end{figure*}

In order to investigate the stability of CO tips regarding lateral manipulation further, we built up iron clusters atom by atom using a CO tip. Figure \ref{fig6}(a) shows a topographic image of five individual iron atoms adsorbed on Cu(111). The constant-height $\Delta f$ image resolves the toroidal structure \cite{Emmrich2015e} of a single adatom (see inset of Fig. \ref{fig6}(a)). Afterwards, the adatoms 1 and 2 were pulled together via lateral manipulation (see topographic image in Fig. \ref{fig6}(b)). The CO tip directly allows to resolve the individual atoms within the formed dimer in $\Delta f$ (see inset Fig. \ref{fig6} (b)). In the topographic image the dimer shows an instability, similar as observed of copper dimers on Cu(111) \cite{Repp2003}, which can be attributed to a switching of the dimer between two or more stable adsorption configurations. However, the $\Delta f$ image does not show an instability. The $\Delta f$ image was acquired at a higher tunneling current setpoint (smaller tip-sample distance) and hence it could be possible that the presence of the tip is stabilizing the dimer in that specific adsorption site observed in the high-resolution $\Delta f$ image. Figures \ref{fig6}(c) through (e) depict the same scan frame of Fig. \ref{fig6}(b) after adding one after the other atom to the cluster. The inset of Fig. \ref{fig6}(e) depicts the internal structure of the created pentamer. Four more pentamers were constructed artificially atom by atom using various CO tips (see Appendix \ref{AppendixF}). All of them arrange in an in-plane geometry which shows a clear preference of the flat cluster over the 3D cluster -- flat and 3D clusters were proposed by DFT claculations in Ref. \cite{Khajetoorians2013a}, while experimentally the cluster geometry could not be resolved in their STM experiments with non-characterized tips.
To prove that the CO tip does not change during the artificial construction of iron clusters from single atoms, $I$- and $\Delta f$-$z$ spectra on \& off a CO molecule before and after the cluster construction, presented in Fig. \ref{fig6}, were acquired (see Fig. \ref{fig7}): The tunneling current image ($z=\mathrm{const.}$) of a single CO molecule adsorbed on the Cu(111) surface after performing the atom-by-atom assembly appear qualitatively similar to the image before creating the iron pentamer and, additionally, the two color bar ranges are identical (see Fig.\ref{fig7}(a) and (b)). Moreover, the distance dependent tunneling current $I\left(z\right)$ spectra, acquired in the center of the CO molecule and above the Cu(111) surface before (light blue curve on CO, red curve on Cu(111)) and after (green curve on CO, black curve on Cu(111)) the pentamer creation, overlay, respectively (see Fig. \ref{fig7}(c)). The residuals $I\left(z\right)_\mathrm{CO,before}-I\left(z\right)_\mathrm{CO,after}$ and $I\left(z\right)_\mathrm{Cu(111),before}-I\left(z\right)_\mathrm{Cu(111),after}$ are within the noise level of the tunneling current amplifier (see Fig. \ref{fig7}(d)). The simultaneously recorded $\Delta f$-$z$-curves above the CO and the Cu(111) surface are also qualitatively identical (see Fig. \ref{fig7}(g)) and the residuals $\Delta f\left(z\right)_\mathrm{Cu(111),before}-\Delta f\left(z\right)_\mathrm{Cu(111),after}$  (blue curve) are within the noise level. The difference $\Delta f\left(z\right)_\mathrm{CO,before}-\Delta f\left(z\right)_\mathrm{CO,after}$  (orange curve) shows a small deviation of less than $0.12\,\mathrm{Hz}$ which can be attributed to slighty different $x$- and $y$-positions of the CO tips on the CO molecule witch respect to each other. The presented data prove that the CO tip did not change during the building process of the pentamer shown in Fig. \ref{fig6}.

Additionally, we conducted sliding experiments with CO tips. Sliding is accessible by reducing the tip-sample distance further such that, the adatom is trapped inside the force field of the tip and following its movements continously, and, hence, differs from the discontinous pulling mode \cite{Bartels1997,Hla2005}. By moving the tip laterally, single iron, copper and silicon adatoms can be slid over the substrate without losing the CO tip. 

We conclude that lateral manipulation of single iron adatoms on Cu(111) with CO tips and monoatomic metal tips occur both in an attractive pulling mode. Furthermore, we find that a slight tilt of the tip causes an asymmetry of the lateral force profiles resembling an atomic ratchet: At a particular tip-sample distance, lateral manipulation of single iron adatoms is only possible in one direction while the direction is opposite using CO tips with respect to monoatomic metal tips. We can explain the reversed influence of tilted CO tips on the asymmetry of the lateral force field by the inverted dipoles of the tips. Moreover, we find that by approaching the CO tip further, single iron, copper and silicon adatom can be slid continuously over the Cu(111) surface. Finally, we show that atom-by-atom assembly of iron clusters via lateral manipulation using CO tips is possible while the high-resolution capability of CO tips can be used for determining their atomic structure. In all experiments, the manipulation was possible without losing the CO from the tip's apex.

\begin{acknowledgments}
The authors thank N. Hauptmann and A. J. Weymouth for fruitful discussions and the Deutsche Forschungsgemeinschaft for funding within the research Projects No. SFB 689, project A9 and CRC 1277, project A02.
\end{acknowledgments}

\appendix

\section{Full set of tunneling current and $\Delta f$ linescans corresponding to the $\Delta f$ linescans in Figs. \ref{fig1}(a) and (b)}
\label{AppendixA}
\begin{figure}
\centering
\includegraphics{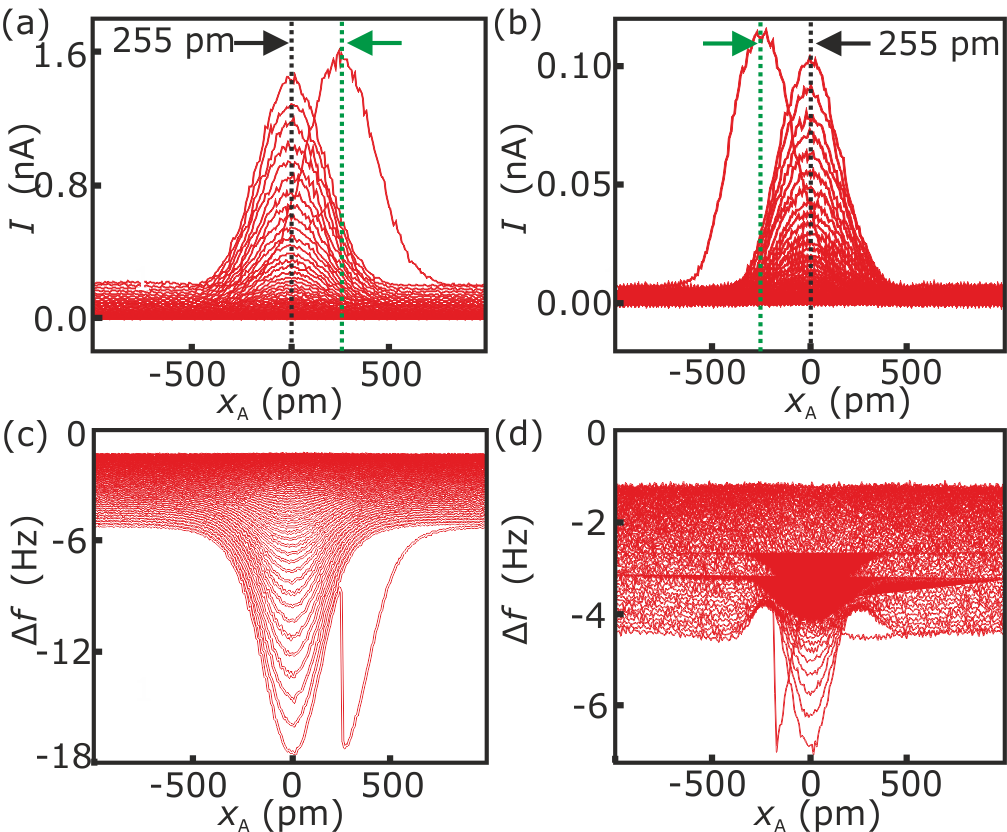}
\caption{Panels (a) and (b) depict the full raw data sets of the tunneling current $I$ in the backward scanning direction simultaneously acquired with the $\Delta f$ linescans in Figs. \ref{fig1}(a) and (b), respectively. A bias voltage of $1\,\mathrm{mV}$ was applied to the tip. In both cases, the maxima of the tunneling current curves after the manipulation $I\left(x,z_\mathrm{man}\right)$ is larger than the previous current curves $I\left(x,z_\mathrm{man}+5\,\mathrm{pm}\right)$ because the tips are closer by $5\,\mathrm{pm}$. Panels (c) and (d) depict the full raw data sets corresponding to the $\Delta f$ data in Figs. \ref{fig1}(a) and (b) along the $+\vec{x}_\mathrm{A}$-direction, respectively.}
\label{figS1}
\end{figure}
Figures \ref{figS1}(a) and (b) show the full set of the tunneling current linescans of the backward direction, resulting from the lateral manipulation experiments of an iron adatom along the $\vec{x}_\mathrm{A}$-direction, corresponding to the frequency shift linescans $\Delta f$ depicted in Figs. \ref{fig1}(a) and (b), respectively. The vertical black dotted line indicates the position of the center of the iron adatom before manipulation ($x_\mathrm{A}=0\,\mathrm{pm}$). Figure \ref{figS1}(a) shows that the iron adatom is manipulated by one atomic position to the next-neighbor fcc hollow site separated by $255\,\mathrm{pm}$ along the positive $\vec{x}_\mathrm{A}$-direction, when using the monoatomic metal tip. In case of the CO tip (Fig. \ref{figS1}(b)), the adatom is manipulated to the next-neighbor fcc hollow site separated by $255\,\mathrm{pm}$ in the $-\vec{x}_\mathrm{A}$-direction. Figures \ref{figS1}(c) and (d) show the full $\Delta f$ data set corresponding to the $\Delta f$ data depicted in Figs. \ref{fig1}(a) and (b).
\section{Error analysis of the absolute height determination with respect to point contact}
\label{AppendixB}
\subsection{Monoatomic metal tips}
\label{AppendixB1}
The tunneling current $I$ in scanning tunneling microscopy is given by $I\left(z\right)=I_0\cdot e^{-2\kappa z}$ while $z$ is the tip-sample distance, $I_0$ is the current at point contact at $z=0$ and $\kappa$ is a decay constant which can be deduced from tunneling current $I$ versus distance $z$ spectra \cite{Binnig1982}. Using the conductance $G\left(z\right)=I\left(z\right)/V=G_0\cdot e^{-2\kappa z}$ with the bias voltage $V$ between tip and sample and the conductance $G_0=2e^2/h=(12906\,\Omega)^{-1}$ for a quantum point contact ($e$ is the elementary charge and $h$ is the Planck constant) \cite{Ternes2008} and assuming that the tunneling current is flowing through a single atom at the tip's apex, an absolute distance of the tip with respect to point contact $G_0$ can be introduced. As described in Ref. \cite{Ternes2008} and in paragraphs S1 and S2 of Ref. \cite{Schneiderbauer2014a}, the assumption that the whole tunneling current only flows through the single atom at the tip's apex introduces an error in determining the absolute tip-sample distance $z$. Comparing the cases of the tunneling current $I_\mathrm{single}$ flowing only through a single atom of the apex of the tip with the tunneling current $I_\mathrm{single+2nd layer}$ flowing through the atom at the tip's apex and $N$ additional atoms in the second layer, spaced by $a/2=361\,\mathrm{pm}/2$ in $z$ (assuming a copper tip in the $\langle 100\rangle$ direction, as proposed by Ref. \cite{Schneiderbauer2014a} in Fig. S1), we can write \cite{Schneiderbauer2014a}:
\begin{equation}
\begin{split}
I_\mathrm{single}\left(z'\right)&=I_\mathrm{single+2nd layer}\left(z\right)\\
I_0\cdot e^{-2\kappa z'}&=I_0\cdot e^{-2\kappa z}+N\cdot I_0\cdot e^{-2\kappa\left(z+a/2\right)}
\end{split}
\end{equation}
When fixing the decay constant $\kappa$ to $10^{10}\,\mathrm{m}^{-1}$, the uncertainty in the determination in the absolute distance to point contact $z-z'$ in dependency of the number of atoms $N$ in the 2nd layer can be calculated (see Table \ref{Table2}).
 \begin{table}
 \caption{Uncertainty $z-z'$ in the determination in the absolute distance to point contact in dependency of the number of atoms $N$ in the 2nd layer.\label{Table2}}
 \begin{ruledtabular}
 \begin{tabular}{l|cccccc}
 $N$&$0$&$4$&$6$&$8$&$10$&$12$ \\
 $z-z'$&$0\,\mathrm{pm}$&$5.1\,\mathrm{pm}$&$7.5\,\mathrm{pm}$&$9.8\,\mathrm{pm}$&$12.0\,\mathrm{pm}$&$14.1\,\mathrm{pm}$\\
 \end{tabular}
 \end{ruledtabular}
 \end{table}
By ignoring the current which is flowing through the second layer the absolute distance to point contact is underestimated by at least $5.1\,\mathrm{pm}$ which is only $10\%$ of the value calculated in paragraph S2 of Ref. \cite{Schneiderbauer2014a}.
\subsection{Monoatomic metal and CO tips}
The conductance at point contact of CO tips $G_0^\mathrm{CO}=(404497\,\Omega)^{-1}$ was determined experimentally \cite{Schneiderbauer2014a}, ignoring current flowing through the monoatomic metal tip behind the CO molecule. Assuming that the tip length is increased by $250\,\mathrm{pm}$ after terminating it with a CO molecule \cite{Schneiderbauer2014a}, a similar calculation as done in section \ref{AppendixB1} can be performed. We compare the underestimation in the tip-sample distance determination for not taking into account the current flowing through the metal tip behind the CO molecule of the tip. The metal tip structure described in section \ref{AppendixB1} is used for this estimation:
\begin{equation}
\begin{split}
G_\mathrm{CO}\left(z'\right)=&G_\mathrm{CO+metal\,tip}\left(z\right)\\
G_0^\mathrm{CO}\cdot e^{-2\kappa z'}=&G_0^\mathrm{CO}\cdot e^{-2\kappa z}+G_0\cdot e^{-2\kappa\left(z+250\,\mathrm{pm}\right)}\\
&+N\cdot G_0\cdot e^{-2\kappa\left(z+250\,\mathrm{pm}+a/2\right)}
\end{split}
\end{equation}
This leads to an underestimation of the tip-sample distance $z$ in dependency of the number of atoms $N$ in the 2nd layer of the monoatomic metal tip behind the CO molecule shown in Table \ref{Table3}.
\begin{table}
 \caption{Underestimation of the tip-sample distance $z$ in dependency of the number of atoms $N$ in the 2nd layer of the monoatomic metal tip behind the CO molecule\label{Table3}}
 \begin{ruledtabular}
 \begin{tabular}{l|cccccc}
 $N$&$0$&$4$&$6$&$8$&$10$&$12$ \\
 $z-z'$&$9.6\,\mathrm{pm}$&$10.5\,\mathrm{pm}$&$11.0\,\mathrm{pm}$&$11.4\,\mathrm{pm}$&$11.9\,\mathrm{pm}$&$12.3\,\mathrm{pm}$\\
 \end{tabular}
 \end{ruledtabular}
 \end{table}
In our experience, the geometrical structure of the metal tip does not change during the pickup of the CO molecule. Therefore, the number of second layer atoms of a metal tip stays the same before and after picking up a CO molecule. Hence, we can calculate the difference in the underestimation of the tip-sample distance of a monoatomic metal tip and a CO-terminated tip depending on the number $N$ of second layer atoms of the metal tip (see Table \ref{Table4}).
\begin{table}
 \caption{Difference in the underestimation of the tip-sample distance of a monoatomic metal tip and a CO-terminated tip depending on the number $N$ of second layer atoms of the metal tip\label{Table4}}
 \begin{ruledtabular}
 \begin{tabular}{l|cccccc}
 $N$&$0$&$4$&$6$&$8$&$10$&$12$ \\
 $z-z'$&$9.6\,\mathrm{pm}$&$5.4\,\mathrm{pm}$&$3.6\,\mathrm{pm}$&$1.6\,\mathrm{pm}$&$0.1\,\mathrm{pm}$&$-1.8\,\mathrm{pm}$\\
 \end{tabular}
 \end{ruledtabular}
 \end{table}
Hence, the absolute distances calibrated with respect to point contact using monoatomic metal tips ($G_0=(12906\,\Omega)^{-1}$) and CO-terminated tips ($G_0^\mathrm{CO}=(404497\,\Omega)^{-1}$) can be compared within an error of less than $5.4\,\mathrm{pm}$ for sufficiently sharp metal tips (assuming $N\geq 4$).

\section{Asymmetric lateral force profiles along the $\vec{x}_\mathrm{B}$-direction for a monoatomic metal tip}
\label{AppendixC}

\begin{figure}
\centering
\includegraphics{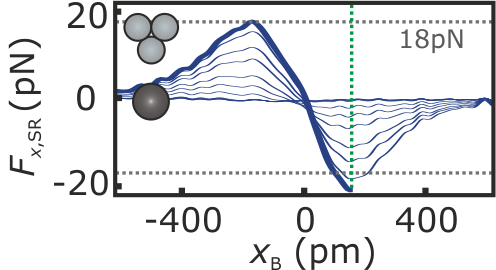}
\caption{Along the manipulation direction $\vec{x}_\mathrm{B}$ in which the monoatomic metal tip shown in Fig. \ref{fig3}(a) is more tilted, the lateral force profiles are more asymmetric with respect to the center of the iron adatom at $x_\mathrm{B}=0\,\mathrm{pm}$ compared to the lateral forces along the $\vec{x}_\mathrm{A}$-direction depicted in Fig. \ref{fig1}(e). In this case, the iron adatom is manipulated after the tip has passed the center of the iron adatom since the lateral forces are higher in absolute value for tip positions $x>0\,\mathrm{pm}$. The green dotted line shows the tip position in the moment of lateral manipulation.}
\label{figS2}
\end{figure}

Figure \ref{figS2} depicts the lateral force acting between a monoatomic metal tip and an iron adatom along the $\vec{x}_\mathrm{B}$-direction. The position $x_\mathrm{B}=0\,\mathrm{pm}$ indicates the center of the iron adatom before manipulation. Compared to Fig. \ref{fig1}(e), the lateral forces are more asymmetric with respect to the center of the adatom. The thick blue curve is the lateral force which corresponds to the tip-sample distance $z_\mathrm{man}$ in which the adatom is manipulated. In this tip-sample height, the maximum force value before the tip has passed the center of the iron adatom ($x_\mathrm{B}<0\,\mathrm{pm}$) is $x_\mathrm{B}=18\,\mathrm{pN}$. This absolute force value is already reached for a tip-sample separation of  $z_\mathrm{man}+5\,\mathrm{pm}$ after the tip has passed the center of the adatom ($x_\mathrm{B}>0\,\mathrm{pm}$), and, hence, it was manipulated after the tip has passed the center of the iron adatom.

\section{Tilt analysis of the COFI images}
\label{AppendixD}
\subsection{Monoatomic metal tip of Fig. \ref{fig3}(a)}
\label{AppendixD1}
\begin{figure}
\centering
\includegraphics{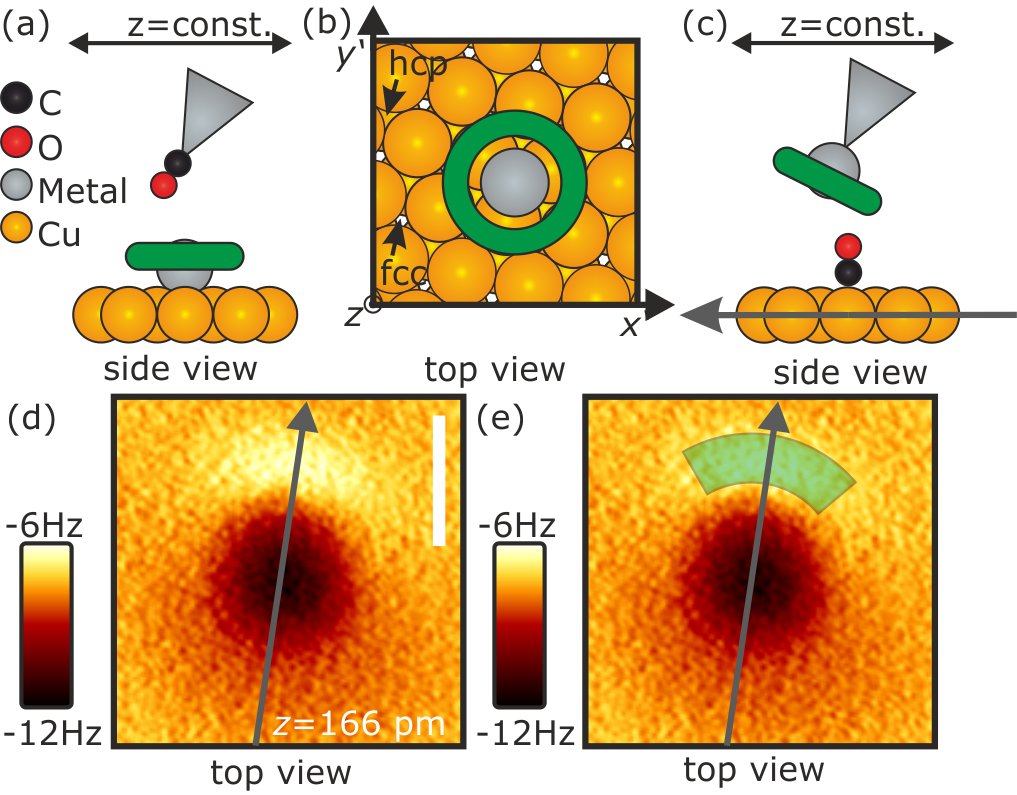}
\caption{Sketch of a measurement of a single metal atom adsorbed on the Cu(111) surface imaged with a CO tip in constant height in (a) side and (b) top-view, respectively. (c) Sketch of a COFI measurement with a tilted monoatomic metal tip. The green torus indicates the repulsive torus resolved in previous experiments \cite{Emmrich2015e}. Panel (d) depicts the COFI image of Fig. \ref{fig3}(a). (e) The repulsive sickle is interpreted as a torus segment of the front most atom of the tip's apex imaged with the CO molecule adsorbed on the surface. The white scale bar resembles a length of $300\,\mathrm{pm}$.}
\label{figS3}
\end{figure}

Imaging single iron and copper adatoms adsorbed on the Cu(111) surface with a CO tip in the close-distance regime results in the appearance of repulsive tori with a single attractive center, while clusters of two and three iron adatoms appear as connected structures of two and three tori, respectively \cite{Emmrich2015e}. Figures \ref{figS3}(a) and (b) depict a sketch of the imaged torus (green color) of a single metal atom adsorbed on the Cu(111) surface, imaged with a CO tip in side and top-view, respectively. Since the tip is scanning the adatom in constant height, the imaged torus is oriented parallel to the surface. In the COFI experiment, the experimental setup is inverted: a CO molecule adsorbed on the Cu(111) surface acts as an effective probe which images the tip which is scanning across the surface \cite{Welker2012,Welker2013} (see Fig. \ref{figS3}(c) for the side view setup). Depending on the amount of atoms terminating the tip's apex, the COFI image will show a single torus for monoatomic metal tips or two- and threefold symmetric structures in case of two and three atoms at the apex of the tip. Figure \ref{figS3}(d) displays the COFI image of the metal tip of Fig. \ref{fig3}(a). It shows a single attractive (dark) center and a repulsive (bright) sickle at the top part of the image which means that the tip is terminated by a single atom. For a non-tilted tip, the repulsive sickle would transform into a torus around the attractive center, similar to the shape of single metal adatoms on the Cu(111) surface imaged by CO tips mentioned above and sketched in Figs. \ref{figS3}(a) and (b). We interpret the observation of the sickle in the top part of the COFI image in Fig. \ref{figS3}(d) as a segment of the torus of the front most atom of the tip. Therefore, we conclude that the tip is tilted along the axis indicated by gray solid arrow, as sketched in Fig. \ref{figS3}(c). This directly yields the COFI image of Fig. \ref{figS3}(d), since the torus is only imaged for positions at the repulsive sickle, indicated by the semi-transparent green torus segment in Fig. \ref{figS3}(e).

\subsection{CO tip of Fig. \ref{fig3}(e)}

\begin{figure*}
\centering
\includegraphics{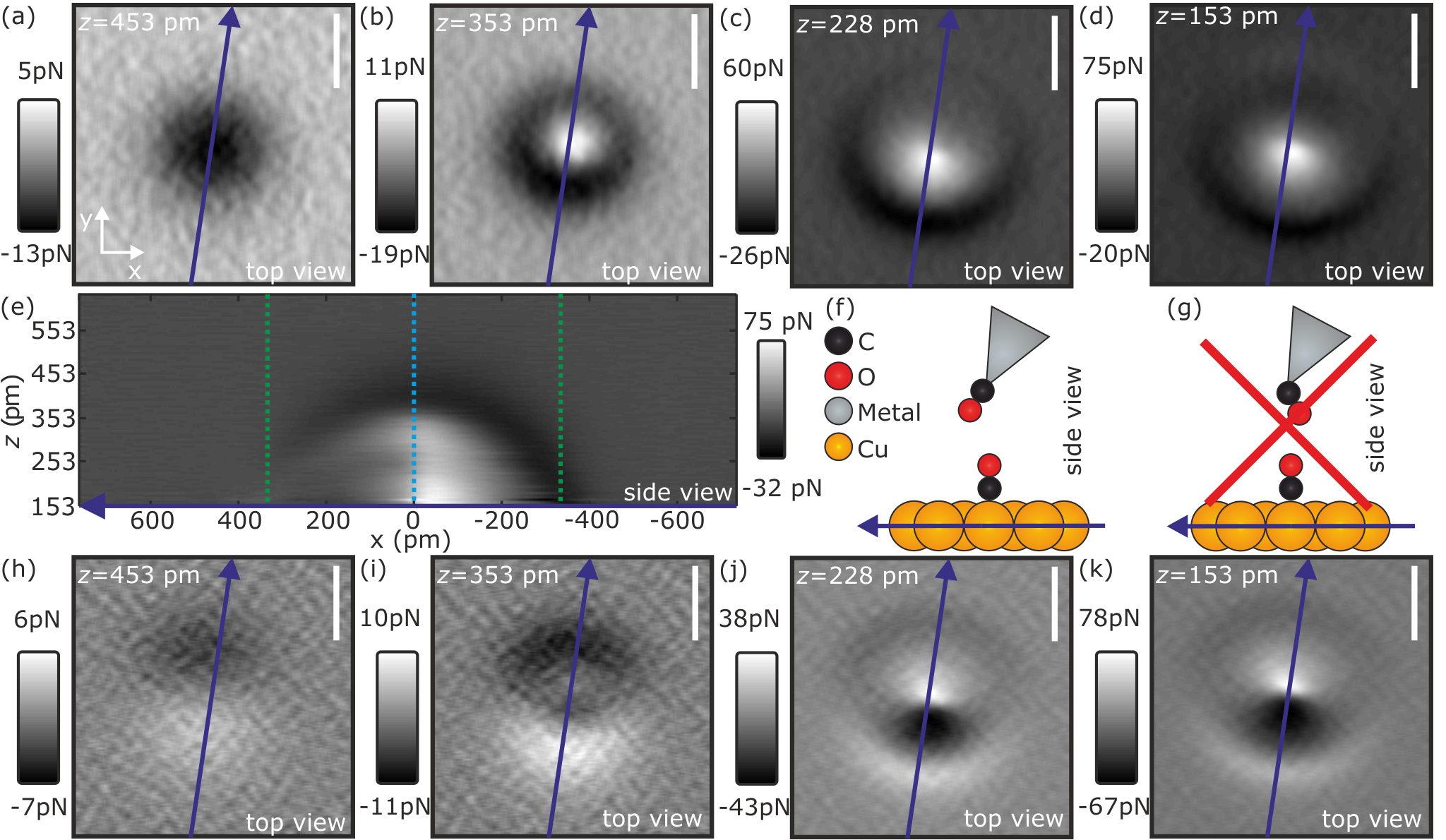}
\caption{Panels (a) through (d) depict the vertical short-range force in four different tip-sample distances. Panel (e) shows the vertical short-range force in the plane defined by the blue arrow and the $z$-axis. Due to symmetry arguments, the adsorption of the CO molecule to the metal tip apex as sketched in panel (f) is more likely than an adsorption sketched in panel (g). Panels (h)-(k) show the short-rang lateral force in the $y$-direction. All $z$ distances are given with respect to point contact $G_0^{\mathrm{CO}}=(404497\,\Omega)^{-1}$ for CO tips \cite{Schneiderbauer2014a}. The scale bars in this figure represent a length of $300\,\mathrm{pm}$.}
\label{figS4}
\end{figure*}	

The monoatomic metal tip presented in Fig. \ref{fig3}(a) and discussed in more detail in section \ref{AppendixD1} was functionalized with a CO molecule and the lateral manipulation experiments were repeated. Figures \ref{figS4}(a) to (d) depict the short-range vertical force of the COFI image of the CO tip for four different tip-sample distances (Fig. \ref{figS4}(c) is the vertical short-range force COFI image corresponding to the $\Delta f$ image shown in Fig. \ref{fig3}(e)). Far away (Fig. \ref{figS4}(a)), the CO appears as an attractive dip. Approaching the tip by $100\,\mathrm{pm}$, the center gets repulsive while an attractive ring around the repulsive center remains (Fig. \ref{figS4}(b)). We note that the center of the remaining attractive ring and the center of the repulsive feature do not match. The center of the attractive ring is shifted along the negative axis indicated by the blue arrow with respect to the center position of the repulsive feature (Fig. \ref{figS4}(b)). A further approach of the tip by twice $75\,\mathrm{pm}$ results in an enlargement of the attractive ring and of the repulsive feature (Fig. \ref{figS4}(c) and (d)), respectively. Figure \ref{figS4}(e) shows a cut of the three-dimensional short-range vertical force dataset through the center of the CO molecule along the blue colored axis. In the side view, the asymmetry of the attractive cap can be observed. For a non-tilted CO tip, the attractive cap (Fig. \ref{figS4}(e)) would be symmetric. Due to the tilt, the cap becomes asymmetric. Figures \ref{figS4}(h)-(k) depict the lateral force along the $y$-direction for the same tip-sample distances as the vertical short-range forces shown in Figs. \ref{figS4}(a)-(d). 
The interpretation of the COFI images of the CO tip is much more complex \cite{Sun2011} compared to the COFI image of the monoatomic metal tip discussed in section \ref{AppendixD1}. Without studying the COFI images in detail, the symmetry argument leads to the same tilt direction of the CO tip as the metal tip as sketched in Fig. \ref{fig3}(f)-(h) and in Fig. \ref{figS4}(f). Due to conservation of the symmetry of the metal tip, the adsorption of the CO on the tip as sketched in Fig. \ref{figS4}(g) is unlikely. This argument is supported by the experimental observation that the COFI image of the CO tip shows an asymmetry along the same axis as the tilt of the monoatomic metal tip (see Fig. \ref{figS4}). This is commonly observed in our experiments. 

\section{Analytical model for lateral manipulation of a CO molecule with a monoatomic metal tip}
\label{AppendixE}

\begin{figure}
\centering
\includegraphics{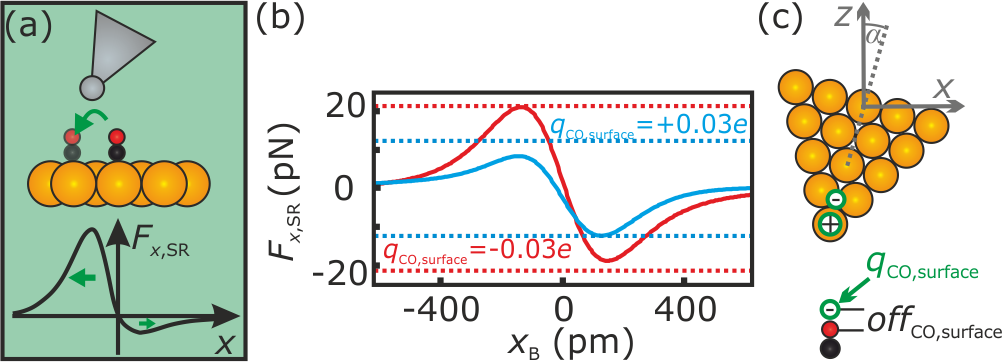}
\caption{(a) The experimentally observed lateral force acting between a monoatomic metal tip and a single CO molecule on the surface is higher in absolute value when the tilted part of the tip is facing the CO molecule \cite{Emmrich2015d}. The analytical model yields reversed asymmetries for a negative charge $q_\mathrm{CO,surface}=-0.03\,e$ (red curve) (a) versus a positive charge $q_\mathrm{CO,surface}=+0.03\,e$ (blue curve) (b) on the oxygen atom. (c) The charges $q_\mathrm{CO,surface}=\pm 0.03\,e$ were placed in a distance of $\textit{off}_\mathrm{CO,surface}=100\,\mathrm{pm}$ with respect to the core of the oxygen atom.}
\label{figS5}
\end{figure}

Emmrich \textit{et al.} \cite{Emmrich2015d} manipulated single CO molecules with monoatomic metal tips. The extracted lateral force curves showed an asymmetry witch respect to the center of the CO molecule. Essentially, it was found that the lateral forces are higher in absolute value when the tilted part of the tip is facing towards the CO molecule as sketched in Fig. \ref{figS5}(a). In the following, the analytical model of the main text will be applied to a monoatomic metal tip scanning a CO molecule on the surface. In the experiments presented in Ref. \cite{Emmrich2015d}, the lateral manipulation took place at a tip distance of about $200\,\mathrm{pm}$ with respect to point contact $G_0$. Taking into account the distance between the core of the oxygen atom to the core of a copper atom of the topmost layer of $302\,\mathrm{pm}$ \cite{Gajdos2005}, the closest core-core-distance between the oxygen atom of the CO molecule on the surface and the tip's apex atom in experiment is given by (see also Fig. \ref{fig4}):
\begin{equation}
200\,\mathrm{pm}+\Delta z_\mathrm{SS}+2r_\mathrm{Cu}-302\,\mathrm{pm}=340\,\mathrm{pm}
\end{equation}
While the direction of the dipole moment of a CO molecule adsorbed to a STM/AFM tip is known \cite{Ellner2016}, the direction of the dipole moment of a CO molecule adsorbed on the surface is under discussion. Schwarz \textit{et al.} \cite{Schwarz2014} report an overall positive dipole of the CO molecule, while Hofmann \textit{et al.} \cite{Hofmann2014a} suggest a negative dipole of the CO adsorbed on the surface. We model both cases with a single charge of $q_\mathrm{CO,surface}=\pm 0.03\,e$ on the CO molecule on the surface (see Fig. \ref{figS5}(b)). The charge is offset, similar to the CO tip charge, by $100\,\mathrm{pm}$ apart from the core of the oxygen atom (see $\textit{off}_\mathrm{CO,surface}$ in Fig. \ref{figS5} (c)). The monoatomic metal tip is modeled in the same way as done above for the iron atom modeling ($q_\mathrm{Cu}=\pm 0.13\,e$, separated by $d_\mathrm{Cu}=135\,\mathrm{pm}$). By comparing the asymmetry of the theoretical lateral forces with the observed asymmetry of the lateral forces in experiment, we find that the negative charge on the oxygen atom of the CO molecule on the surface qualitatively reproduces the experimentally observed asymmetry (compare Figs. \ref{figS5}(a) and (b)). Hence, by considering a negative charge on the oxygen atom on the surface, the analytical model describes the asymmetric lateral force curves observed in Ref. \cite{Emmrich2015d} with a monoatomic metal tip on CO/Cu(111).

\section{Simultaneous $I$ and $\Delta f$ images of five individually created pentamers using lateral manipulation with CO tips}
\label{AppendixF}
\begin{figure*}
\centering
\includegraphics{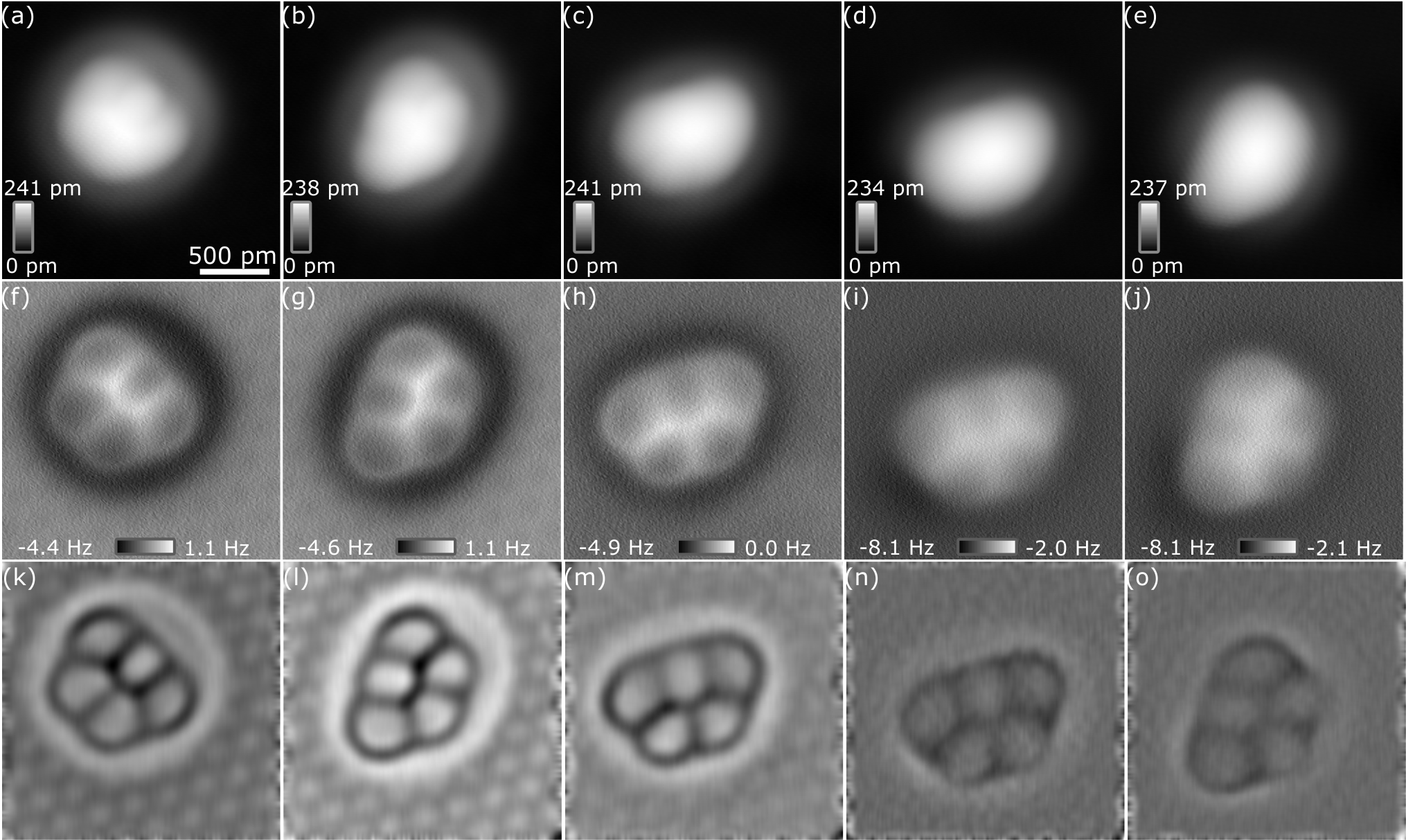}
\caption{Panels (a) through (e) depict the topographic images of five different pentamers, which were all created by controlled lateral manipulation using CO tips. The two pentamers shown in panels (a) and (b) were created and imaged with the same CO tip. The pentamer depicted in panel (c) was built up and imaged with another CO tip. The pentamers depicted in panel (d) and (e) were created and imaged with a third CO tip. Panels (f)-(j) depict the simultaneously acquired frequency shift images which resolve the individual atoms within the clusters. Panels (k) through (o) show the frequency shift images of panels (f) through (j) after post processing in order to enhance the contrast (Gaussian filter ($\sigma=10\,\mathrm{pm}$), Laplace filter, Gaussian filter ($\sigma=10\,\mathrm{pm}$)). All observed pentamers adsorb in an in-plane geometry. In all images of this figure, the tunneling current setpoint was set to $300\,\mathrm{pA}$, except of the cluster shown in the middle column where a tunneling current setpoint of $400\,\mathrm{pA}$ was used. The bias voltage was set to $-10\,\mathrm{mV}$ and all images consist of 256 pixel by 256 pixel.}
\label{figS6}
\end{figure*}

\end{document}